\PassOptionsToPackage{cmyk,table}{xcolor}
\documentclass[conference]{iaria} 
\pdfoutput=1 

\usepackage[babel=true,english=american]{csquotes}
\usepackage[USenglish]{babel}
\usepackage{amsmath}
\usepackage{paralist}
\usepackage[caption=false,font=footnotesize]{subfig}
\usepackage[
  babel=true, 
  expansion=alltext,
  protrusion=alltext-nott, 
  nopatch=eqnum, 
  final 
]{microtype}

\title{Towards Unsupervised Adversarial Document Detection in Retrieval Augmented Generation Systems}

\author{
    \IEEEauthorblockN{Patrick Levi\,\orcidlink{0000-0002-5216-4555} 
    }
    \IEEEauthorblockA{%
        Department of Electrical Engineering, Media, and Computer Science\\
        Ostbayerische Technische Hochschule Amberg-Weiden\\
        Amberg, Germany\\
        e-mail: {\tt p.levi@oth-aw.de}
    }
}

\usepackage[nolist]{acronym}

\begin{acronym}
\acro{llm}[LLM]{Large Language Model}
\acro{rag}[RAG]{Retrieval Augmented Generation}
\end{acronym}

\usepackage{float}
\usepackage{relsize}
\usepackage{booktabs}
\usepackage{longtable} 
\usepackage{caption}

\begin{document}
    
\maketitle

\begin{abstract}
Retrieval augmented generation systems have become an integral part of everyday life. Whether in internet search engines, email systems, or service chatbots, these systems are based on context retrieval and answer generation with large language models. With their spread, also the security vulnerabilities increase. Attackers become increasingly focused on these systems and various hacking approaches are developed. Manipulating the context documents is a way to persist attacks and make them affect all users. Therefore, detecting compromised, adversarial context documents early is crucial for security. While supervised approaches require a large amount of labeled adversarial contexts, we propose an unsupervised approach, being able to detect also zero day attacks. We conduct a preliminary study to show appropriate indicators for adversarial contexts. For that purpose generator activations, output embeddings, and an entropy-based uncertainty measure turn out as suitable, complementary quantities. With an elementary statistical outlier detection, we propose and compare their detection abilities. Furthermore, we show that the target prompt, which the attacker wants to manipulate, is not required for a successful detection. Moreover, our results indicate that a simple context summary generation might even be superior in finding manipulated contexts. 

\end{abstract}

\section{Introduction}
Over the last few years, Retrieval Augmented Generation (RAG) systems  \cite{2020-rag-overview} have become a valuable and indispensable tool in our lives, in particular as a support for knowledge intensive activities. Almost every current chatbot is conceptualized as a RAG system, ranging from internet search engines over email bots to customer service chatbots. 
RAG systems combine the user query (prompt) with a document database, which contains potential supporting information. A retriever component finds documents that provide information regarding the prompt (context), while the generator Large Language Model (LLM) creates an answer to the prompt based on these contexts. Thus, RAG systems efficiently combine the power of LLMs with large information collections. 
Due to their powers, they also attract attackers, offering potentially high returns: The information database may contain private or confidential information which can be extract, the RAG system can be tricked into spreading biased or false information, or it can be sabotaged into shutdown.
To reach their goal, hackers could attack the prompt directly, but usually they sneak Adversarial Contexts (ACs) into the context document database. This way, the attack becomes persistent and potentially effects all users. Often, the context database is at least partly accessible to attackers, e.g., for email bots: An attacker can add a document by sending an email to the victim.
Therefore, ACs are a serious security threat for RAG systems.%
While defenses exist, these defenses might imply a huge effort and thus a downgrade of the system usability (e.g., human in the loop). Adversarial mechanism are basically those to jailbreak LLMs in general \cite{jailbreakingliu23, levi-iaria2024}, which are continuously adapted and improved. Therefore, the best strategy against ACs is detecting them as early as possible and preventing the attack entirely. Detection has to account for continuously new (zero day) attacks. Many approaches exist, to differentiate between adversarial and non-adversarial documents, e.g., leveraging the power of supervised learning \cite{2025poisonedrag}. However, supervised learning requires large amounts of labeled documents and usually only detects attacks similar to those learned from the data. These effects are well known from network Intrusion Detection (ID) \cite{2024-quintian-comparison}. Similar to ID applications, we want to switch from supervised to unsupervised approaches, designing an Adversarial Context Detector (ACD), which is flexible, robust and does not require knowledge about the specific attack type, nor much labeled AC examples.

The development of our approach is currently work in progress, however, we present a promising initial study answering two main questions. First, which quantities indicate ACs reliably enough for an ACD. Second, is it necessary to know the target question (prompt), which the attacker intended to manipulate. We use a simple statistical outlier detection to answer these questions, thus showing the overall feasibility of unsupervised ACD. Our paper is organized as follows: We summarize related work in Section \ref{sec:related}, present our statistical approach in Section \ref{sec:method}, and propose experiments on an adversarial dataset in Section \ref{sec:experiments}. We evaluate our results on this dataset in Section \ref{sec:results}. In Section \ref{sec:conclusions} we conclude on the feasibility of ACD with our indicator variables and outline future work.

\section{Related Work}\label{sec:related}
Various attacks on RAG system have been investigated recently by researchers and ethical hackers alike. Some attacks target private or confidential information in the context documents by either performing membership inference attacks \cite{2025-rag-mia-privacy} or even extract whole documents from the database \cite{2024-rag-privacy-attacks} using adversarial prompts. Other approaches use adversarial documents to cause the RAG system to reveal whole documents. In the case of an email agent with the permission to send emails, this has been shown to extend even into a self-replicating worm-like attack \cite{2025-rag-ai-worm}. Attackers who want the RAG system to spread false information usually add poisoned context documents \cite{zhong2023poisoning}. Dedicated context poisoning can also trick safety alignment in LLMs to cause refusing answers, thus generating a denial-of-service attack \cite{2025-rag-dos-jamming}, which is not easy to detect.
The techniques used in the prompt as well as in adversarial documents are the same as for LLM jailbreaking \cite{2025poisonedrag}. ACs are usually created, using oracle based text manipulation, targeted whitebox optimizations \cite{zou-arro} or heuristic optimization methods \cite{2024garag}. Often, attacks need a smart combination of tricking the retriever, as well as the generator components to achieve their goal efficiently \cite{2024-badrag}.
Recently, various works have been published to detect adversarial attacks against LLMs. Layer activations have been shown to be successful indicators for adversarial attacks and even be useful for classification of attack types \cite{ball_understanding_2024}. Smart supervised approaches have been applied to detect adversarial attacks, dealing with the problem of few examples and small amounts of labeled data \cite{2025-prevrag}.\\
We plan to extend this research to unsupervised approaches using anomaly detection to find adversarial contexts, as has been done in adversarial image detection \cite{2024-ad-images}.

\section{Method}\label{sec:method} 
Our ACD feasibility study is based on the following threat model. The hacker intends to poison a specific question or a group of questions with targeted adversarial documents. The defender is not aware of the target question, but needs to detect the attack by screening the contexts. The attacker can add context documents, which is realistic for common applications using internet contexts or emails. These contexts might be found by the retriever and thus find their way into the generator's context. 
The defender must detect the attack only based on the contexts. We assume they have three potential knowledge levels: They can query the RAG generator arbitrarily and have access to the text output, eventually also to the corresponding logits, or even to the generator LLM layer activations. We assess the capability of our ACD depending on these knowledge levels.\\
Generator activations are used directly as an adversarial attack indicator, where we restrict ourselves to the last layer activations. From the logits, we compute the TokenSAR score \cite{2024-TokenSAR}, an entropy based measure \cite{2024-steindl-entropy}, which has been proven to be successful for jailbreak detection \cite{2026-eacl}. 
The generator output text is encoded into a 768-dimensional embedding vectors using a variant of the MPNet model \cite{2020-mpnet, mpnet-all-hf}. We chose this model since it is small and thus cheap to deploy and operate.\\
To realize our ACD, we want to work without the target prompt. Therfore, we use a simple summary prompt asking the generator model to summarize available context information (''Summarize the following context documents: <contexts>. Consider every important aspect in your summary.''). We evaluate activations, TokenSAR, and embeddings for the related answer. 
We consider these quantities obtained for $N$ combinations of non-adversarial contexts as a reference and compute the mean value and standard deviation for the $N$ TokenSAR values. For the $N$ activations and embeddings, respectively, we first compute their center and the Euclidean distance of each individual vector to it. Then, we consider the mean distance and its standard deviation. We repeat the summary prompt, replacing at least one context with an AC, and extract the same quantities from the generator LLM outputs. For activations and embeddings, we consider Euclidean distances from the previously computed center. 
Subsequently, we apply Grubb's test at $\alpha=0.1$ to determine whether any of the quantities $q_\text{adv}$ (TokenSAR, embeddings, activations) obtained with the ACs is an outlier with respect to the reference values:
\begin{align}
    q_\text{adv}\notin& [\mu -G_{\text{crit}} ;\mu +G_{\text{crit}} ]\\
    G_{\text{crit}} &=\frac{N-1}{\sqrt{N}}\sqrt{\frac{t^2}{N-+t^2}}
\end{align}
with $\mu ,s$ being the empirical mean and standard deviation of the reference values and $t = t_{1 - \frac{\alpha}{2N},\, N-2}$ being the two-sided quantile of Student's t-distribution.
We implicitly assume normal distribution of the quantities, being aware of the interdependence due to the center distances. This way, we have a semi-supervised method, requiring only valid contexts to obtain the reference values.

\section{Experiments} \label{sec:experiments}
\begin{figure}[h!]
    \centering%
    \includegraphics[page=1,scale=0.22]{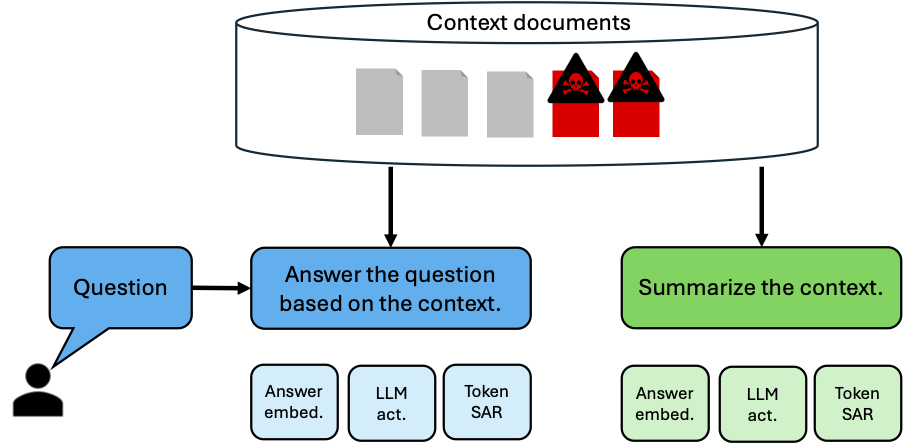}
	\caption{Schematic diagram of our approach: The question prompt requires the user question and the contexts, while the simpler summary prompt only requires the contexts.}
	\label{fig:approach}
\end{figure}

Our experiments are performed based on 100 questions from the HotpotQA dataset \cite{2018-hotpotqa} with adversarial contexts from PoisonedRAG \cite{2025poisonedrag, poisonedrag-data}. Every HotpotQA question consists of, among other things, a question, related context documents from Wikipedia, and a correct answer. PoisonedRAG adds adversarial contexts and a target incorrect answer. For each question in the dataset, we conduct a small study using $N=10$ combinations of valid contexts. To simulate the retriever component, we select $k=5$ valid contexts from the original set. The RAG generator is realized with a Llama-3.1 8B model \cite{2024-llama3}. Using the summary prompt introduced in Section \ref{sec:method}, each context combination is summarized. The output is processed to obtain its embeddings, TokenSAR values and the activations (see Figure \ref{fig:approach}). In addition, the generator LLM is prompted with the question and the contexts (question prompt: ''Answer the following question based on the provided context: Question: <question>. Contexts: <contexts>.'') to verify that the contexts are valid and informative enough for the generator to be able to answer the question correctly. The answer is verified to be semantically equivalent to the documented correct answer from the dataset using Mistral-7B \cite{jiang2023mistral7b} as a judge. If the answer cannot be verified as correct, it is discarded, leading eventually to less than $N=10$ valid contexts. In case there are too few valid contexts, the question is discarded.

In the same way as for the summaries, also the answers to the question prompt are processed to obtain embeddings, TokenSAR values and the last layer activations are extracted. A comparison between question and summary will show the impact of not knowing the target prompt.\\
Afterwards, each question is evaluated using the same prompts but introducing AC to the contexts. The $k=5$ valid contexts are successively replaced by AC, starting with the first one until more than half the contexts are AC. This way, we cover various degrees of manipulations in our study.

\section{Results and Discussion} \label{sec:results}
\subsection{Results} 
We ran one experiment on HotpotQA dataset, using both, the summary prompt as well as the question prompt. Comparing both results we will see the effect of knowing the targeted question or instruction. 
Table \ref{tab:quant-results} summarizes how many cases using at least one AC can be detected by our ACD. There are up to 30  such context combinations per question containing at least one AC. 

\begin{table}[htpb!]
  \centering
  \caption{Quantitative results for the different indicators and prompt types (question or summary)}  \label{tab:quant-results}
  \begin{center}
  \begin{tabular}{|l||c|c|}
  \hline
  \textbf{Quantity}&\textbf{Summary}&\textbf{Question}\\ \hline\hline
  Invalid &11& 11 \\ \hline
  Undetected &486& 486 \\ \hline
  \textit{Detected by} &&\\
  TokenSAR + Emb. only&60 &60 \\ 
  TokenSAR + Act. only&38&22 \\ 
  TokenSAR + Emb. + Act. &1912 &1928 \\ \hline
  TokenSAR total&2010 &2010\\
  Emb. total &1972&1988 \\
  Act. total &1950&1950 \\ \hline
  \end{tabular}
  \end{center}
\end{table}

Questions which did not reveal a sufficient number of correct answers for valid contexts are marked as invalid. For all others we count, how many AC cases are exclusively detected by any combination of our indicators. Moreover, we count how often each indicator is successful. The number of detected AC cases were counted, independent of attack success.\\
TokenSARs turn out to have the highest predictive capability, followed by activations and embeddings. Clearly, the summary is equally suited for ACD as the question prompt. None of the indicator appears to be dominant over the others. Every attack is detectable by at least two indicators, most attacks even by all three of them. For both prompts, about 19.5\;\% (486 of 2496) of attacks remain undetected.

\subsection{Discussion}
The results are promising towards ACD for two reasons: First, it turned out that it is not necessary for the defender to know the question the attacker tried to manipulate. Using a summary prompt turned out sufficient to detect AC. We interpret this as follows: When we increased the number of AC, we conclude that that with few AC contradictions to valid contexts might be found in our indicators, while for increasing number of AC, the summary is different enough to be an outlier compared to the valid context summary. These effects could well be reflected in the (last layer) activations of the generator. Further, they influence the summary text, since a specific statement either does not occur in it or it is weakened to resolve the controversy. In the TokenSAR, AC are visible due to a higher uncertainty in the answer logits of the generator LLM, resulting in a higher entropy.\\
Second, our results show that several indicators are able to detect manipulations. Embeddings and TokenSAR values are the most successful ones. While activations require no extra computation overhead, they might not be available, especially for proprietary models and APIs. Since the TokenSARs are based on the output logits, they are easy to compute and often available. Output embeddings are always available. The computational effort can be reduced by using a small embedding model.

\section{Conclusion and Future Work}\label{sec:conclusions}
Detection of adversarial attacks against RAG systems is highly relevant for the future of chatbots and other systems. Attackers have two attack paths, the prompt and the context documents. The latter are especially exposed in important applications, like email bots. Attacks against the context documents need to be reliably prevented, since they lead to persistent misbehavior of the system. Therefore, reliable ACD methods are needed. In contrast to existing supervised approaches, we conducted a preliminary study to look into unsupervised detection methods. Using a simple statistical approach, we successfully identified three indicator quantities that are well suited for ACD: the generator LLM activations, output embeddings, and entropy-based TokenSAR. Systematic comparison of these indicators revealed that they show a high redundancy, but also complement each other in some cases. Furthermore, we found that ACD can be realized without knowing the specific prompt (question, instruction, etc.) that is targeted by the attacker. A simple summary instruction is sufficient. Therefore, the goal of this small, preliminary study, to elucidate promising approaches for unsupervised ACD has been achieved. Future research shall focus on the following key aspects. First, increase the experimental runs to obtain more normal contexts, while this study used a relatively small number. With more samples, a proper quantitative evaluation of false positives (valid contexts incorrectly detected as AC) will be possible, too. Furthermore, the approach needs to be extended to more RAG attacks beyond PosionedRAG \cite{2025poisonedrag}. In particular, corpus poisoning \cite{zhong2023poisoning}, BADRAG \cite{2024-badrag}, and GARAG \cite{2024garag} shall be considered. 
Finally, more sophisticated anomaly detection methods shall be used, following previous work from network intrusion detection \cite{2024-quintian-comparison}.

\begingroup
\sloppy
\printbibliography[notcategory=selfref]

@misc{jailbreakingliu23,
      title={Jailbreaking {C}hat{G}{P}{T} via Prompt Engineering: An Empirical Study}, 
      author={Yi Liu and Gelei Deng and Zhengzi Xu and Yuekang Li and Yaowen Zheng and Ying Zhang and Lida Zhao and Tianwei Zhang and Yang Liu},
      year={2023},
      eprint={2305.13860},
      archivePrefix={arXiv},
      primaryClass={cs.SE},
      volume={arXiv:2305.13860},
      note={version 1},
      doi={10.48550/arXiv.2305.13860}
}

@misc{zou-arro,
title={Universal and Transferable Adversarial Attacks on Aligned Language Models}, 
author={Andy Zou and Zifan Wang and Nicholas Carlini and Milad Nasr and J. Zico Kolter and Matt Fredrikson},
year={2023},
eprint={2307.15043},
archivePrefix={arXiv},
primaryClass={cs.CL},
  doi = {https://doi.org/10.48550/arXiv.2307.15043},
  url = {https://arxiv.org/abs/2307.15043},
  %publisher = {arXiv},
  %volume={arXiv:2307.15043},
  note={version 2}
}

@misc{2024garag,
      title={Typos that Broke the {RAG}'s Back: Genetic Attack on {RAG} Pipeline by Simulating Documents in the Wild via Low-level Perturbations}, 
      author={Sukmin Cho and Soyeong Jeong and Jeongyeon Seo and Taeho Hwang and Jong C. Park},
      year={2024},
      eprint={2404.13948},
      archivePrefix={arXiv},
      primaryClass={cs.CL},
      url={https://arxiv.org/abs/2404.13948}, 
doi = {https://doi.org/10.48550/arXiv.2404.13948},
publisher = {arXiv},
volume={arXiv:2404.13948},
note={version 1}
}

@misc{2024-badrag,
      title={BadRAG: Identifying Vulnerabilities in Retrieval Augmented Generation of Large Language Models}, 
      author={Jiaqi Xue and Mengxin Zheng and Yebowen Hu and Fei Liu and Xun Chen and Qian Lou},
      year={2024},
      eprint={2406.00083},
      archivePrefix={arXiv},
      primaryClass={cs.CR},
      url={https://arxiv.org/abs/2406.00083}, 
doi = {https://doi.org/10.48550/arXiv.2406.00083},
publisher = {arXiv},
volume={arXiv:2406.00083},
note={version 2},
}

@inproceedings{2020-rag-overview,
author = {Lewis, Patrick and Perez, Ethan and Piktus, Aleksandra and Petroni, Fabio and Karpukhin, Vladimir and Goyal, Naman and K\"{u}ttler, Heinrich and Lewis, Mike and Yih, Wen-tau and Rockt\"{a}schel, Tim and Riedel, Sebastian and Kiela, Douwe},
title = {Retrieval-augmented generation for knowledge-intensive NLP tasks},
year = {2020},
isbn = {9781713829546},
publisher = {Curran Associates Inc.},
address = {Red Hook, NY, USA},
booktitle = {Proceedings of the 34th International Conference on Neural Information Processing Systems},
articleno = {793},
numpages = {16},
location = {Vancouver, BC, Canada},
series = {NIPS '20}
}

@inproceedings{2024-steindl-entropy,
    title = "Linguistic Obfuscation Attacks and Large Language Model Uncertainty",
    author = {Steindl, Sebastian  and
      Sch{\"a}fer, Ulrich  and
      Ludwig, Bernd  and
      Levi, Patrick},
    editor = {V{\'a}zquez, Ra{\'u}l  and
      Celikkanat, Hande  and
      Ulmer, Dennis  and
      Tiedemann, J{\"o}rg  and
      Swayamdipta, Swabha  and
      Aziz, Wilker  and
      Plank, Barbara  and
      Baan, Joris  and
      de Marneffe, Marie-Catherine},
    booktitle = "Proceedings of the 1st Workshop on Uncertainty-Aware NLP (UncertaiNLP 2024)",
    month = mar,
    year = "2024",
    address = "St Julians, Malta",
    publisher = "Association for Computational Linguistics",
    url = "https://aclanthology.org/2024.uncertainlp-1.4",
    pages = "35--40",
}

@inproceedings{2018-hotpotqa,
  title={{HotpotQA}: A Dataset for Diverse, Explainable Multi-hop Question Answering},
  author={Yang, Zhilin and Qi, Peng and Zhang, Saizheng and Bengio, Yoshua and Cohen, William W. and Salakhutdinov, Ruslan and Manning, Christopher D.},
  booktitle={Conference on Empirical Methods in Natural Language Processing ({EMNLP})},
  year={2018}
}

@inproceedings{2026-eacl,
  author={Rubenbauer, Franziska and Steindl, Sebastian and Levi, Patrick and Loebenberger, Daniel and Sch{\"a}fer, Ulrich},
  title={Detection of Adversarial Prompts with Model Predictive Entropy},
  year = {2026},
  publisher = {Association for Computational Linguistics},
  booktitle = {Findings of the Association for Computational Linguistics: EACL 2026},
  address   = {Rabat, Morocco},
  publisher = {Association for Computational Linguistics},
  note      = {Accepted for publication}
}

@inproceedings{2025poisonedrag,
author = {Zou, Wei and Geng, Runpeng and Wang, Binghui and Jia, Jinyuan},
title = {{PoisonedRAG}: knowledge corruption attacks to retrieval-augmented generation of large language models},
year = {2025},
%isbn = {978-1-939133-52-6},
publisher = {USENIX Association},
address = {USA},
booktitle = {Proceedings of the 34th {USENIX} Conference on Security Symposium},
articleno = {197},
numpages = {18},
location = {Seattle, WA, USA},
series = {SEC '25}
}

@inproceedings{zhong2023poisoning,
   title={Poisoning Retrieval Corpora by Injecting Adversarial Passages},
   author={Zhong, Zexuan and Huang, Ziqing and Wettig, Alexander and Chen, Danqi},
   booktitle={Empirical Methods in Natural Language Processing (EMNLP)},
   year={2023}
}

@inproceedings{2025-prevrag,
    title = "{R}ev{PRAG}: Revealing Poisoning Attacks in Retrieval-Augmented Generation through {LLM} Activation Analysis",
    author = "Tan, Xue  and
      Luan, Hao  and
      Luo, Mingyu  and
      Sun, Xiaoyan  and
      Chen, Ping  and
      Dai, Jun",
    editor = "Christodoulopoulos, Christos  and
      Chakraborty, Tanmoy  and
      Rose, Carolyn  and
      Peng, Violet",
    booktitle = "Findings of the Association for Computational Linguistics: {EMNLP} 2025",
    month = nov,
    year = "2025",
    address = "Suzhou, China",
    publisher = "Association for Computational Linguistics",
    url = "https://aclanthology.org/2025.findings-emnlp.698/",
    doi = "10.18653/v1/2025.findings-emnlp.698",
    pages = "12999--13011",
    %ISBN = "979-8-89176-335-7"
}

@inproceedings{2025-rag-ai-worm,
author = {Cohen, Stav and Bitton, Ron and Nassi, Ben},
title = {Here Comes the {AI} Worm: Preventing the Propagation of Adversarial Self-Replicating Prompts Within {GenAI} Ecosystems},
year = {2025},
%isbn = {9798400715259},
publisher = {Association for Computing Machinery},
address = {New York, NY, USA},
url = {https://doi.org/10.1145/3719027.3765196},
doi = {10.1145/3719027.3765196},
booktitle = {Proceedings of the 2025 ACM SIGSAC Conference on Computer and Communications Security},
pages = {3975–3989},
numpages = {15},
keywords = {adversarial machine learning, ai security, rag, worms},
location = {Taipei, Taiwan},
series = {CCS '25}
}

@inproceedings{2025-rag-mia-privacy,
    title = "Safeguarding Privacy of Retrieval Data against Membership Inference Attacks: Is This Query Too Close to Home?",
    author = "Choi, Yujin  and
      Park, Youngjoo  and
      Byun, Junyoung  and
      Lee, Jaewook  and
      Park, Jinseong",
    editor = "Christodoulopoulos, Christos  and
      Chakraborty, Tanmoy  and
      Rose, Carolyn  and
      Peng, Violet",
    booktitle = "Findings of the Association for Computational Linguistics: EMNLP 2025",
    month = nov,
    year = "2025",
    address = "Suzhou, China",
    publisher = "Association for Computational Linguistics",
    url = "https://aclanthology.org/2025.findings-emnlp.438/",
    doi = "10.18653/v1/2025.findings-emnlp.438",
    pages = "8241--8258",
    %ISBN = "979-8-89176-335-7"
}

@misc{2025-rag-dos-jamming,
      title={Machine Against the {RAG}: Jamming Retrieval-Augmented Generation with Blocker Documents}, 
      author={Avital Shafran and Roei Schuster and Vitaly Shmatikov},
      year={2025},
      eprint={2406.05870},
      note={version 4},
      archivePrefix={arXiv},
      primaryClass={cs.CR},
      url={https://arxiv.org/abs/2406.05870}, 
}

@inproceedings{2024-rag-privacy-attacks,
    title = "The Good and The Bad: Exploring Privacy Issues in Retrieval-Augmented Generation ({RAG})",
    author = "Zeng, Shenglai  and
      Zhang, Jiankun  and
      He, Pengfei  and
      Liu, Yiding  and
      Xing, Yue  and
      Xu, Han  and
      Ren, Jie  and
      Chang, Yi  and
      Wang, Shuaiqiang  and
      Yin, Dawei  and
      Tang, Jiliang",
    editor = "Ku, Lun-Wei  and
      Martins, Andre  and
      Srikumar, Vivek",
    booktitle = "Findings of the Association for Computational Linguistics: ACL 2024",
    month = aug,
    year = "2024",
    address = "Bangkok, Thailand",
    publisher = "Association for Computational Linguistics",
    url = "https://aclanthology.org/2024.findings-acl.267/",
    doi = "10.18653/v1/2024.findings-acl.267",
    pages = "4505--4524"
}

@article{levi-iaria2024,
author = {Levi, Patrick and Neumann, Christoph},
year = {2024},
month = {12},
pages = {214-225},
title = {Goal Hijacking Using Adversarial Vocabulary for Attacking Vulnerabilities of Large Language Model Applications},
volume = {17},
doi = {10.5281/zenodo.14680185}
}

@misc{ball_understanding_2024,
	title = {Understanding {Jailbreak} {Success}: {A} {Study} of {Latent} {Space} {Dynamics} in {Large} {Language} {Models}},
	shorttitle = {Understanding {Jailbreak} {Success}},
	url = {http://arxiv.org/abs/2406.09289},
	doi = {10.48550/arXiv.2406.09289},
	language = {en},
	%publisher = {arXiv},
  archivePrefix={arXiv},
	author = {Ball, Sarah and Kreuter, Frauke and Panickssery, Nina},
	month = oct,
	year = {2024},
}

@inproceedings{2024-TokenSAR,
    title = "Shifting Attention to Relevance: Towards the Predictive Uncertainty Quantification of Free-Form Large Language Models",
    author = "Duan, Jinhao  and
      Cheng, Hao  and
      Wang, Shiqi  and
      Zavalny, Alex  and
      Wang, Chenan  and
      Xu, Renjing  and
      Kailkhura, Bhavya  and
      Xu, Kaidi",
    editor = "Ku, Lun-Wei  and
      Martins, Andre  and
      Srikumar, Vivek",
    booktitle = "Proceedings of the 62nd Annual Meeting of the Association for Computational Linguistics (Volume 1: Long Papers)",
    month = aug,
    year = "2024",
    address = "Bangkok, Thailand",
    publisher = "Association for Computational Linguistics",
    url = "https://aclanthology.org/2024.acl-long.276/",
    doi = "10.18653/v1/2024.acl-long.276",
    pages = "5050--5063"
}

@misc{jiang2023mistral7b,
      title={Mistral 7{B}}, 
      author={Albert Q. Jiang and Alexandre Sablayrolles and Arthur Mensch and Chris Bamford and Devendra Singh Chaplot and Diego de las Casas and Florian Bressand and Gianna Lengyel and Guillaume Lample and Lucile Saulnier and Lélio Renard Lavaud and Marie-Anne Lachaux and Pierre Stock and Teven Le Scao and Thibaut Lavril and Thomas Wang and Timothée Lacroix and William El Sayed},
      year={2023},
      note={version 1},
      eprint={2310.06825},
      archivePrefix={arXiv},
      primaryClass={cs.CL},
      url={https://arxiv.org/abs/2310.06825}, 
}

@article{2024-ad-images,
	author = {Hui Liu and Bo Zhao and Jiabao Guo and Kehuan Zhang and Peng Liu},
	doi = {https://doi.org/10.1016/j.patcog.2023.110127},
	issn = {0031-3203},
	journal = {Pattern Recognition},
	keywords = {Deep neural networks, Adversarial examples, Adversarial detection, Autoencoder, Isolation forest},
	pages = {110127},
	title = {A lightweight unsupervised adversarial detector based on autoencoder and isolation forest},
	url = {https://www.sciencedirect.com/science/article/pii/S0031320323008245},
	volume = {147},
	year = {2024},
	%bdsk-url-1 = {https://www.sciencedirect.com/science/article/pii/S0031320323008245},
	%bdsk-url-2 = {https://doi.org/10.1016/j.patcog.2023.110127}
}

@misc{mpnet-all-hf,
  author={Huggingface},
  title={all-mpnet-base-v2},
  url = {https://huggingface.co/sentence-transformers/all-mpnet-base-v2},
  note={2026.03.10}
}

@misc{poisonedrag-data,
  author={Zou, Wei and Geng, Runpeng and Wang, Binghui and Jia, Jinyuan},
  title={{PoisonedRAG}},
  url={https://github.com/sleeepeer/PoisonedRAG/blob/main/results/adv_targeted_results/hotpotqa.json},
  note={2026.03.10}
}

@misc{2020-mpnet,
      title={MPNet: Masked and Permuted Pre-training for Language Understanding}, 
      author={Kaitao Song and Xu Tan and Tao Qin and Jianfeng Lu and Tie-Yan Liu},
      year={2020},
      note={version 2},
      eprint={2004.09297},
      archivePrefix={arXiv},
      primaryClass={cs.CL},
      url={https://arxiv.org/abs/2004.09297}, 
}

@incollection{2024-quintian-comparison,
	address = {Cham},
	title = {A {Comparison} of {AI}-{Enabled} {Techniques} for the {Detection} of {Attacks} in {IoT} {Devices}},
	volume = {957},
	%isbn = {978-3-031-75015-1 978-3-031-75016-8},
	url = {https://link.springer.com/10.1007/978-3-031-75016-8_21},
	language = {en},
	booktitle = {International {Joint} {Conferences}},
	publisher = {Springer Nature Switzerland},
	author = {Cabeza-Lopez, Eduardo Manuel and Ruiz-Gonzalez, Ruben and Merino-Gomez, Alejandro and Curiel-Herrera, Leticia Elena and Rincon, Jaime Andres},
	editor = {Quintián, Héctor and Corchado, Emilio and Troncoso Lora, Alicia and Pérez García, Hilde and Jove, Esteban and Calvo Rolle, José Luis and Martínez De Pisón, Francisco Javier and García Bringas, Pablo and Martínez Álvarez, Francisco and Herrero Cosío, Álvaro and Fosci, Paolo},
	year = {2024},
	doi = {10.1007/978-3-031-75016-8_21},
	note = {Series Title: Lecture Notes in Networks and Systems},
	keywords = {abstract good},
	pages = {227--236},
}
\endgroup 

\end{document}